\documentclass{aa}
\usepackage[T1]{fontenc} % Uses T1 font encoding.
\usepackage{graphicx}
\usepackage{natbib} 
\usepackage{url}
\usepackage{xcolor}
\usepackage{amsmath}
\usepackage{multirow} 
\usepackage{xy} 
\xyoption{all} 

\newcommand{\soft}[1]{\texttt{#1}}
\newcommand{\GILDAS}{\soft{GILDAS}}
\newcommand{\CLASS}{\soft{CLASS}}
\newcommand{\WEEDS}{\soft{WEEDS}}

\newcommand{\emm}[1]{\ensuremath{#1}}
\newcommand{\emr}[1]{\emm{\mathrm{#1}}}
\newcommand{\unit}[1]{\emr{\,#1}}
\newcommand{\chem}[1]{\ensuremath{\mathrm{#1}}}
\newcommand{\paren}[1]{\emm{\left(  #1 \right)  }} % Parenthesis
\newcommand{\bracket}[1]{\emm{\left[  #1 \right] }} % Brackets

\newcommand{\ps}{\unit{s^{-1}}}

\newcommand{\pscm}{\unit{cm^{-2}}}
\newcommand{\mm}{\unit{mm}}
\newcommand{\mim}{\unit{\mu m}}
\newcommand{\kms}{\unit{km\,s^{-1}}}
\newcommand{\K}{\unit{K}}
\newcommand{\mK}{\unit{mK}}

\newcommand{\mKkms}{\unit{mK\,km\,s^{-1}}}
\newcommand{\Hz}{\unit{Hz}}
\newcommand{\kHz}{\unit{kHz}}
\newcommand{\MHz}{\unit{MHz}}
\newcommand{\GHz}{\unit{GHz}}
\newcommand{\Debye}{\unit{Debye}}

\renewcommand{\ion}[2]{\mbox{#1{\sc #2}}}
\newcommand{\HII}{\ion{H}{ii}} % HII

\renewcommand{\H}{\chem{H}}
\newcommand{\HH}{\chem{H_{2}}}
\newcommand{\HCOp}{\chem{HCO^{+}}}
\newcommand{\HthCOp}{\chem{H^{13}CO^{+}}}
\newcommand{\DCOp}{\chem{DCO^{+}}}
\newcommand{\CFp}{\chem{CF^{+}}}
\newcommand{\Cp}{\chem{C^+}}

\newcommand{\CCHp}{\chem{C_{2}H^+}}
\newcommand{\CCCHp}{\chem{C_3H^{+}}}
\newcommand{\CCCHHp}{\chem{C_{3}H_{2}^{+}}}
\newcommand{\CCCHHHp}{\chem{C_{3}H_{3}^{+}}}

\newcommand{\CCH}{\chem{C_{2}H}}
\newcommand{\CCHH}{\chem{C_{2}H_{2}}}
\newcommand{\CCCH}{\chem{C_{3}H}}
\newcommand{\cCCCH}{c-\chem{C_{3}H}}
\newcommand{\lCCCH}{l-\chem{C_{3}H}}
\newcommand{\lCCCHp}{l-\chem{C_{3}H^+}}
\newcommand{\CCCHH}{\chem{C_{3}H_2}}
\newcommand{\cCCCHH}{c-\chem{C_{3}H_2}}
\newcommand{\lCCCHH}{l-\chem{C_{3}H_2}}
\newcommand{\CCCCH}{\chem{C_{4}H}}

\newcommand{\Tpeak}{\emm{T_\emr{peak}}}

\newcommand{\nH}{\emm{n_\emr{H}}}

\newcommand{\Fsou}{\emm{\nu_\emr{sou}}}
\newcommand{\Flsr}{\emm{\nu_\emr{lsr}}}
\newcommand{\vlsr}{\emm{v_\emr{lsr}}}
\newcommand{\clight}{\emm{c}}

\newcommand{\J}{\emm{J}}
\newcommand{\Be}{\emm{B}}
\newcommand{\De}{\emm{D}}
\newcommand{\He}{\emm{H}}

\newcommand{\A}[2]{\emm{A_{#1 \rightarrow #2}}}
\newcommand{\Av}{\emr{A_v}} % Visual Extinction in Magnitude Units
\newcommand{\magn}{\emr{\,mag}}

\newcommand{\likelihood}{\emm{\mathcal{L}}}

\newcommand{\cf} {cf.}
\newcommand{\ie} {{\em i.e.}}
\newcommand{\eg} {{\em e.g.}}
\newcommand{\etal} {et al.}

\newcommand{\FigSpec}{%
  \begin{figure*}
    \centering{} %
    \includegraphics[width=\hsize]{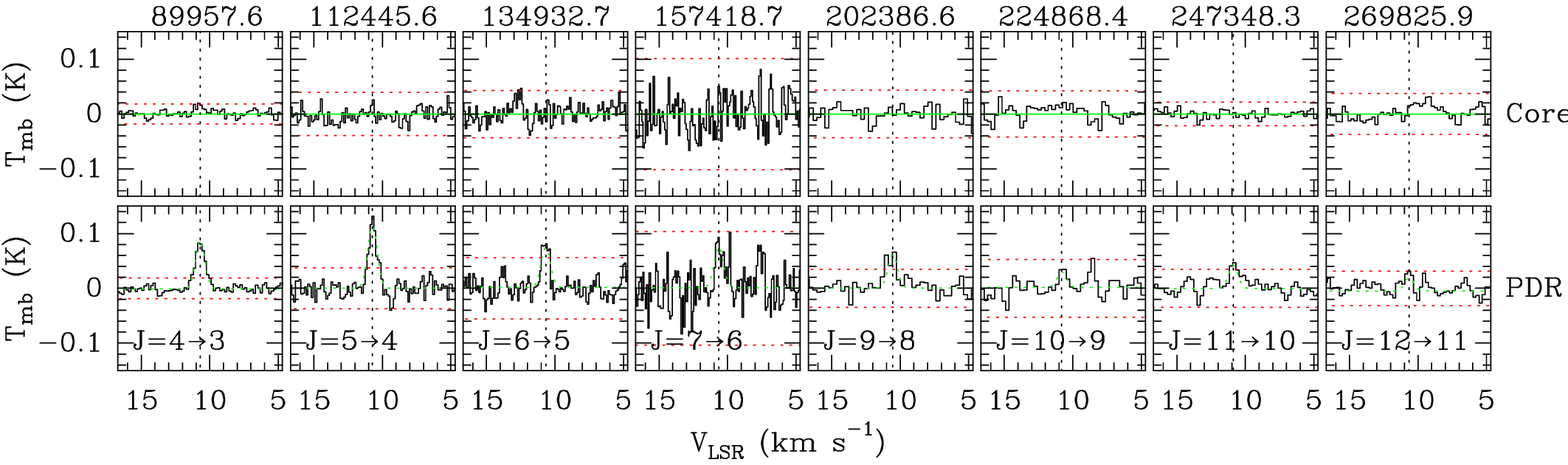}
    \caption{Millimeter lines attributed to the \lCCCHp{} cation in the PDR
      (lower panels) and the dense core (upper panels) positions. The
      numbers above each panel are the line rest frequencies in MHz. The
      spectra and the Gaussian fits are shown as black histograms and green
      curves, respectively. The inferred rest frequencies are displayed as
      vertical dotted black lines for $\vlsr = 10.7\kms$ and $\pm3\sigma$
      noise levels as horizontal dotted red lines.}
    \label{fig:c3hp:spec}
  \end{figure*}}

\newcommand{\FigHydrocarbonsPdBI}{%
  \begin{figure*}
    \centering{} %
    \includegraphics[width=\hsize]{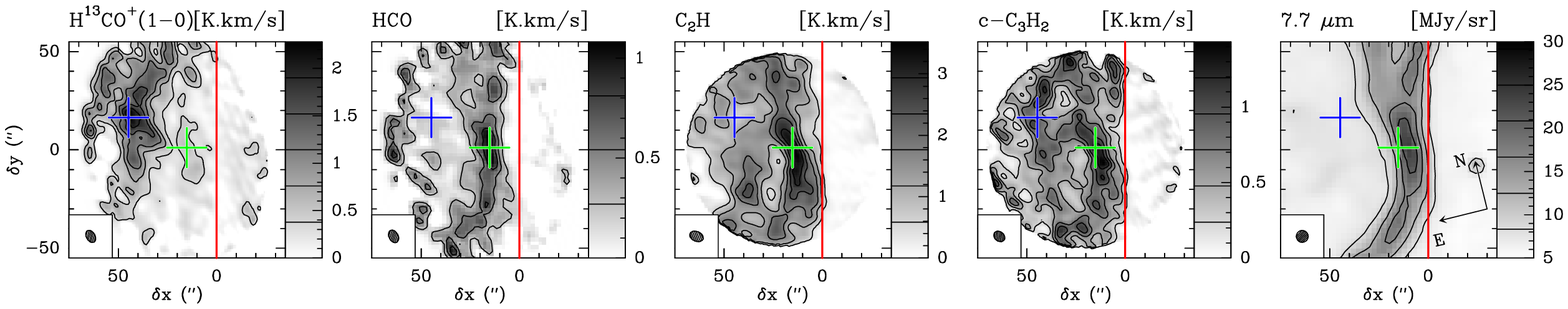}
    \caption{High angular resolution maps of the integrated intensity 
      of \HthCOp{}, HCO, \CCH{}, and \cCCCHH{}, and the 7.7\mim{} PAH
      emission. Maps are rotated by 14$^\circ$ counter-clockwise around the
      projection center, located at $(\delta x, \delta y) = (20'', 0'')$,
      to bring the illuminated star direction in the horizontal direction.
      The horizontal zero is set at the PDR edge.  The emission of all
      lines is integrated between 10.1 and 11.1\kms{}.  Displayed
      integrated intensities are expressed in the main beam temperature
      scale. Contour levels are shown in the grey-scale lookup tables.  The
      red vertical line shows the PDR edge and the blue and green crosses
      show the dense core and PDR positions, respectively.}
    \label{fig:hydrocarbons:pdbi} 
  \end{figure*}}
  
\newcommand{\FigProfile}{%
  \begin{figure}
    \centering{} %
    \includegraphics[width=\hsize]{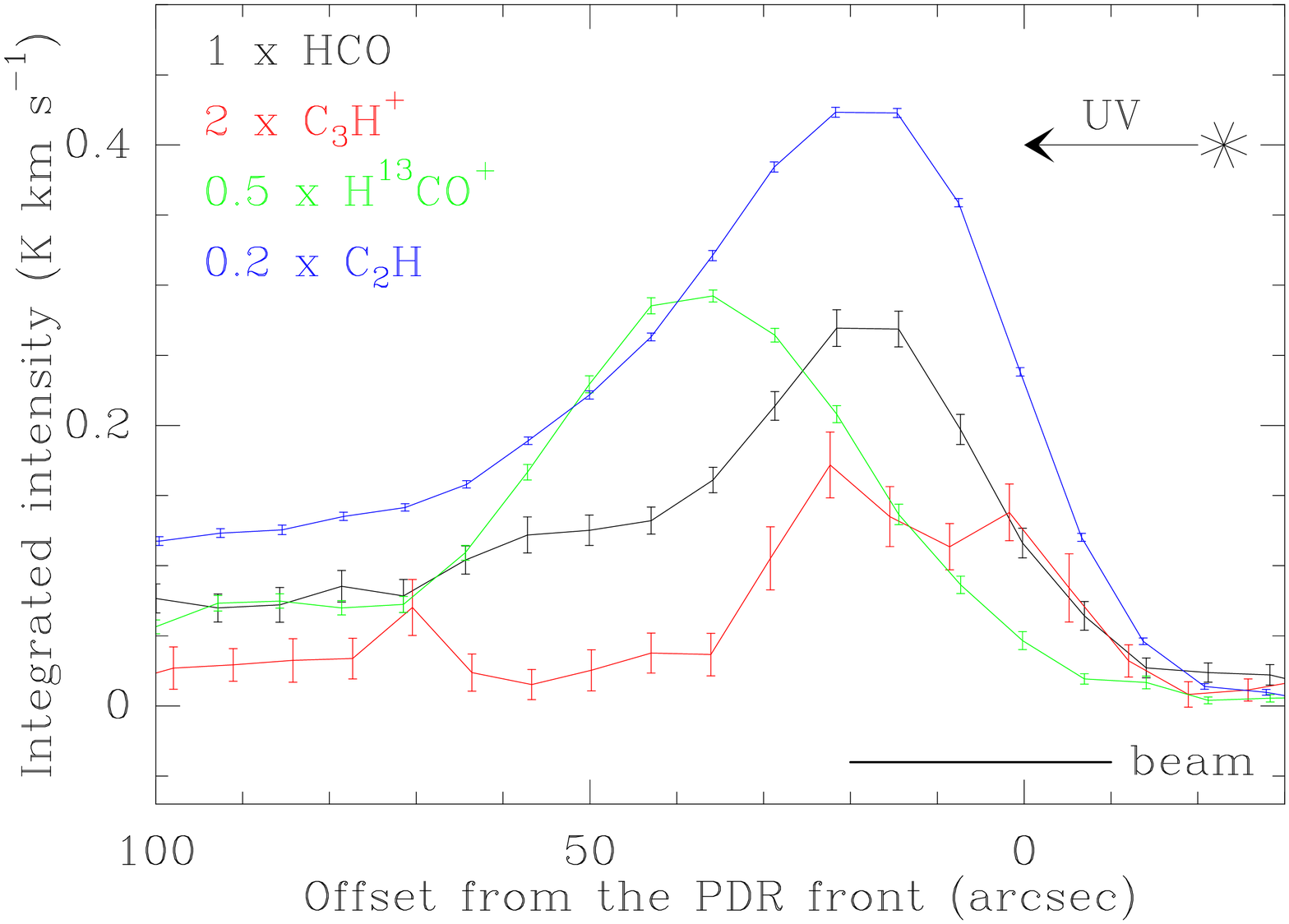}
    \caption{Integrated intensity profiles across the PDR
      photo-dissociation front of the \lCCCHp{}, HCO, \HthCOp{}, and \CCH{}
      species (IRAM-30m). The cut would appear horizontal at $\delta y =
      0''$ on Fig.~\ref{fig:hydrocarbons:pdbi}. The observed lines are
      summarized in Table~\ref{tab:profile}.}
    \label{fig:profile} 
  \end{figure}}

\newcommand{\FigRotDiag}{%
  \begin{figure}
    \centering{} %
    \includegraphics[width=\hsize]{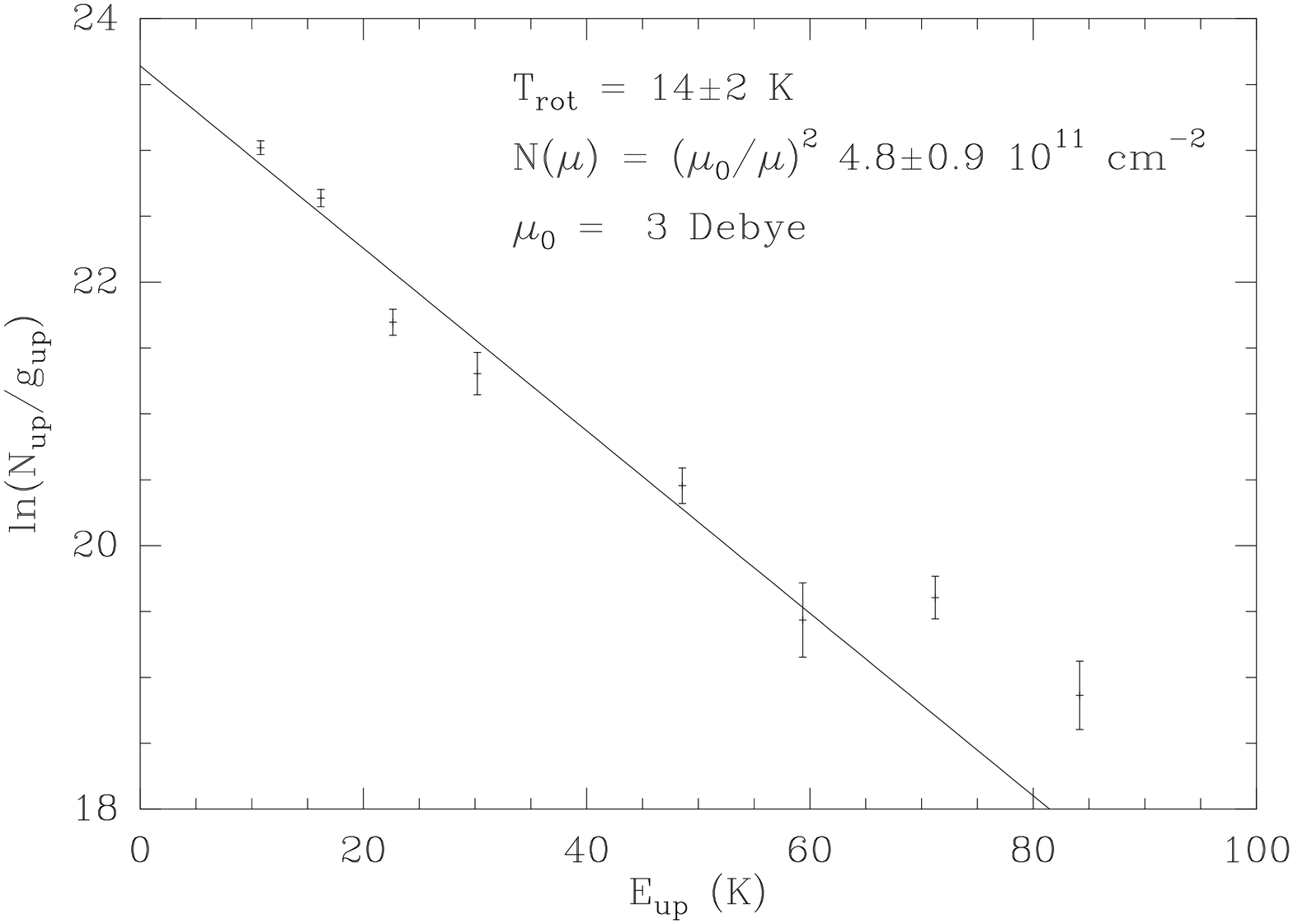}
    \caption{Rotational diagram of \lCCCHp{} at the PDR position.}
    \label{fig:rot:diag} 
  \end{figure}}

\newcommand{\FigChemistryLog}{%
  \begin{figure*}
    \centering{} %
    \includegraphics[width=0.8\hsize]{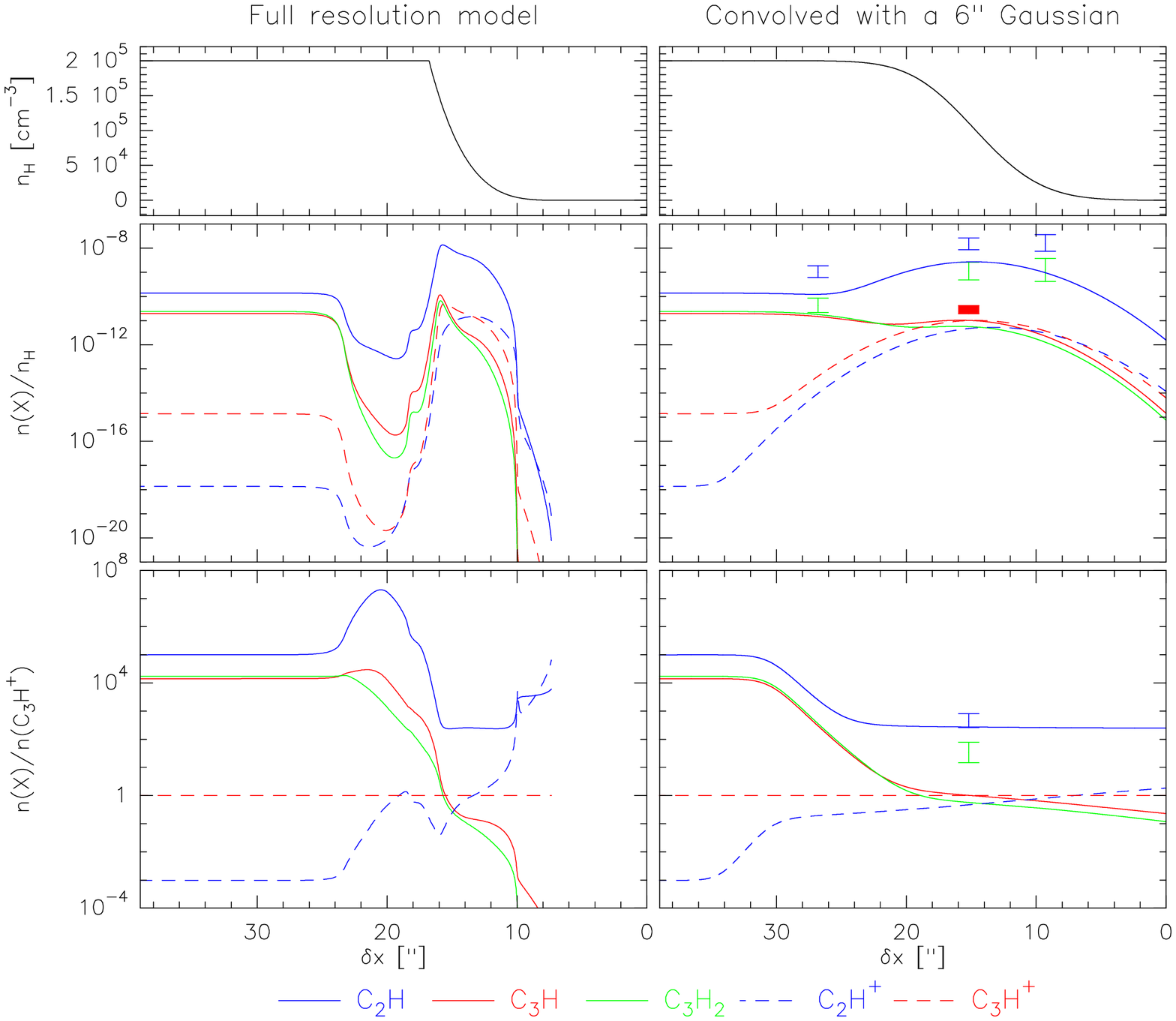}
    \caption{Photochemical model of the Horsehead PDR displayed at 
      the full resolution of the model \textbf{(left column)} and convolved
      with a Gaussian of $6''$-FWHM \textbf{(right column)}. \textbf{Top:}
      Horsehead density profile $\nH = n(\H) + 2n(\HH)$. \textbf{Middle:}
      Predicted abundance of selected small hydrocarbons molecules and
      cations.  \textbf{Bottom:} Abundances of the same hydrocarbons
      relative to the abundance of \CCCHp{}. The illuminating star is
      positioned at the right of the plots. The symbols present the
      measured range of possible abundances for \CCH{} (blue vertical
      segment), \CCCHH{} (green vertical segment), and \CCCHp{} (red filled
      rectangle), inferred at a typical resolution of $6''$. The legend at
      the bottom of the figure presents the line coding for the curves:
      \CCH{} in plain blue, \CCHp{} in dashed blue, \CCCH{} in plain red,
      \CCCHp{} in dashed red, and \CCCHH{} in green.}
    \label{fig:chemistry:log} 
  \end{figure*}}

\newcommand{\FigChemistryLin}{%
  \begin{figure}
    \centering{} %
    \includegraphics[width=\hsize]{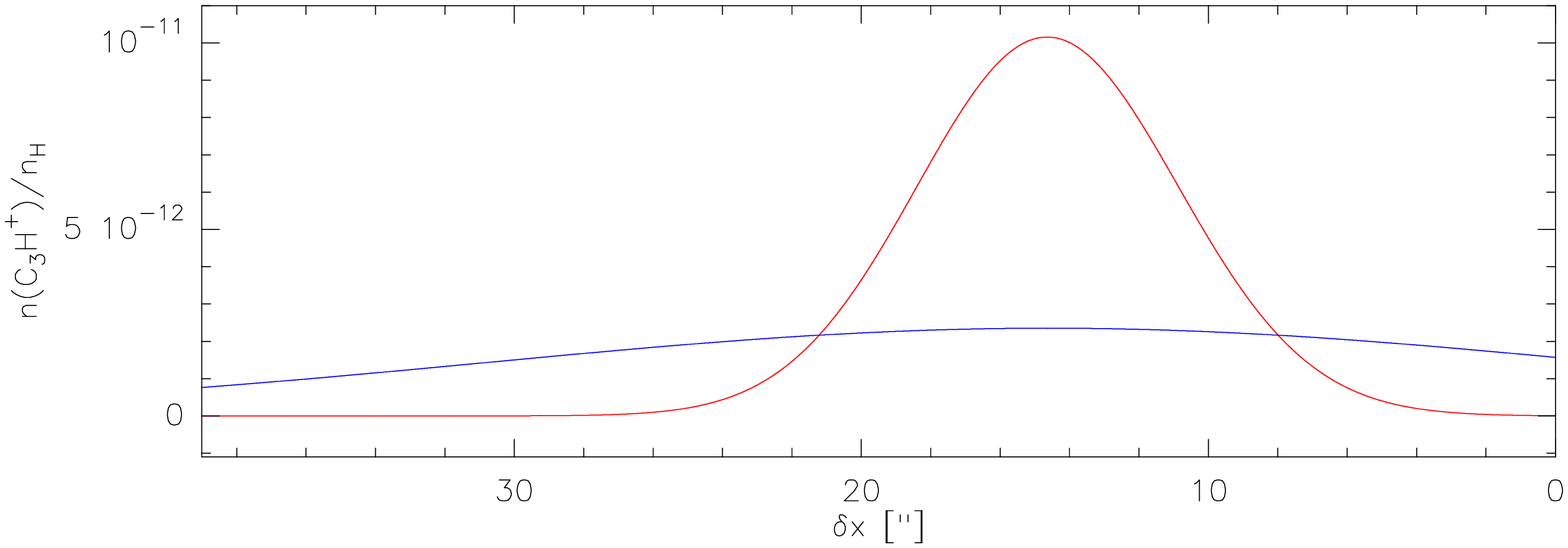}
    \caption{Spatial profiles of the predicted abundance of \CCCHp{} cation.
      The modeled abundance profile was convolved along the $x$-axis with a
      Gaussian of $6''$-FWHM (red line), and $27''$-FWHM (blue line).}
    \label{fig:chemistry:lin} 
  \end{figure}}

\newcommand{\FigHydrocarbonsSD}{%
  \begin{figure*}
    \centering{} %
    \includegraphics[width=\hsize]{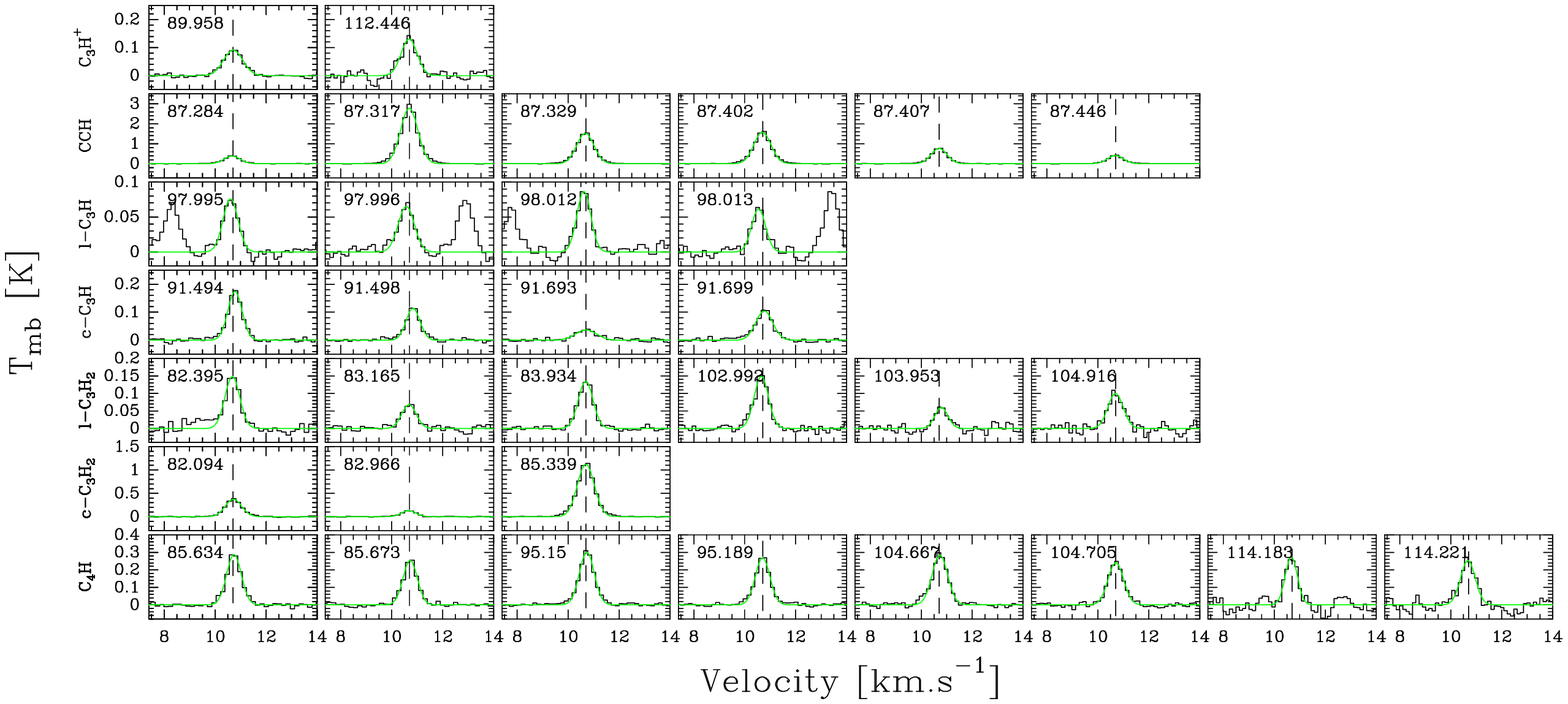}
    \caption{Hydrocarbon lines detected toward the Horsehead PDR 
      at RA=5h40m53.936s, Dec=$-2^\circ28'00''$ (J2000).  Only lines in the
      3\mm{} band, whose energy of the upper level is lower than 100\K{}
      and which have a higher signal-to-noise ratio than 8, are shown.  The
      green lines are Gaussian fits. The vertical dashed lines mark
      10.7~$\kms$.}
    \label{fig:hydrocarbons:30m} 
  \end{figure*}}

\newcommand{\TabProfile}{%
  \begin{table}
    \centering{}
    \caption{Lines simultaneously observed in the cut displayed in 
      Fig.~\ref{fig:profile}.}
    \tiny{%
    \begin{tabular}{ccc}
      \hline
      \hline
      Species   & Transition & Frequency \\
                &            & MHz       \\
      \hline
      \HthCOp{} & $J = 1-0$                     & 86754.2884\\
      HCO       & $1_{0,1},1/2,1-0_{0,0},1/2,1$ & 86777.4600\\
      \CCH{}    & $N=1-0,J=3/2-1/2,F=2-1$       & 87316.8980\\
      \lCCCHp{} & $J = 4-3$                     & 89957.6250\\
      \hline
    \end{tabular}}
    \label{tab:profile}
  \end{table}}

\newcommand{\TablCCCHp}{%
  \begin{table*}
    \centering{} %
    \caption{Results for the simultaneous fit of the \lCCCHp{} 
      line parameters (bottom) and the associated spectroscopic 
      modeling (top).}
    \begin{tabular}{ccccc}
      \hline
      \hline
      Order & $\ln(\likelihood)^{a}$ & Parameter$^{b}$ & Value & Unit \\  
      \hline 
      Second & 2797.9 & \Be{}  & $11244.9474 \pm 0.0007$ & MHz \\
             &        & \De{}  & $     7.652 \pm 0.011$  & kHz \\
      \hline
      Third  & 2802.5 & \Be{}  & $11244.9512 \pm 0.0015$ & MHz \\
             &        & \De{}  & $     7.766 \pm 0.040$  & kHz \\
             &        & \He{}  & $     0.56  \pm 0.19$   &  Hz \\
      \hline
    \end{tabular}
    \smallskip{} %
    {\tiny %
      \begin{tabular}{lrrrcccccccc}
        \hline
        \hline          
        Transition &  Frequency & \multicolumn{2}{c}{Resolutions}  & Offset$^{c}$ & RMS noise$^{c}$ & \Tpeak{}$^{c}$ &  $W^{c}$      & SNR & $A_{ul}$ & $E_u$ & $g_u$ \\
                   &  \MHz{}    &         arcsec & \kHz{}          & \mK{}        & \mK{}           & \mK{}          &  \mKkms{}     &     &  \ps{}   &  \K{} &       \\
        \hline
        $J=  4 \rightarrow  3 $ & $ 89957.625 \pm 0.004 $ & 27.3 &  49 & $-0.8 \pm 0.5$ & $ 5.8 \pm 0.4$ & $ 89$ & $77 \pm  4$ & 19 & $3.5\times10^{-5}$ & 10.8 &  9\\ 
        $J=  5 \rightarrow  4 $ & $112445.642 \pm 0.005 $ & 21.9 &  49 & $+0.2 \pm 1.0$ & $13.9 \pm 0.7$ & $115$ & $99 \pm  7$ & 15 & $6.8\times10^{-5}$ & 16.2 & 11\\ 
        $J=  6 \rightarrow  5 $ & $134932.733 \pm 0.010 $ & 18.2 &  49 & $+1.7 \pm 1.2$ & $17.6 \pm 0.9$ & $ 72$ & $62 \pm  7$ & 10 & $1.2\times10^{-4}$ & 22.7 & 13\\ 
        $J=  7 \rightarrow  6 $ & $157418.719 \pm 0.016 $ & 15.6 &  49 & $+1.3 \pm 2.1$ & $33.2 \pm 1.5$ & $ 75$ & $65 \pm 11$ &  6 & $1.9\times10^{-4}$ & 30.2 & 15\\ 
        $J=  8 \rightarrow  7 $ & $179903.429 \pm 0.024 $ &    d & --- &            --- &            --- &   --- &         --- &--- & $2.9\times10^{-4}$ & 38.8 & 17\\ 
        $J=  9 \rightarrow  8 $ & $202386.678 \pm 0.029 $ & 12.2 & 195 & $+0.6 \pm 1.5$ & $13.1 \pm 1.0$ & $ 62$ & $54 \pm  8$ &  7 & $4.1\times10^{-4}$ & 48.6 & 19\\ 
        $J= 10 \rightarrow  9 $ & $224868.302 \pm 0.033 $ & 10.9 & 195 & $+1.7 \pm 1.7$ & $15.6 \pm 1.2$ & $ 30$ & $26 \pm  9$ &  3 & $5.7\times10^{-4}$ & 59.4 & 21\\ 
        $J= 11 \rightarrow 10 $ & $247348.134 \pm 0.046 $ &  9.9 & 195 & $+0.0 \pm 1.2$ & $11.5 \pm 0.8$ & $ 45$ & $39 \pm  6$ &  6 & $7.6\times10^{-4}$ & 71.2 & 23\\ 
        $J= 12 \rightarrow 11 $ & $269826.003 \pm 0.082 $ &  9.1 & 195 & $-5.3 \pm 1.3$ & $12.3 \pm 0.9$ & $ 25$ & $21 \pm  7$ &  3 & $9.9\times10^{-4}$ & 84.2 & 25\\ 
        \hline
      \end{tabular}}
    $^{a}$ \likelihood{} is the fit likelihood. $^{b}$ Parameters for a linear rotor
    developed to second and third orders (see Sect.~\ref{sec:obs}). $^{c}$
    Results of the simultaneous Gaussian fits using the third-order model 
    to predict the frequencies. The common line width is $0.81 \pm
    0.03\kms$. \\
    $^{d}$ Outside of the EMIR receiver tuning range.
    \label{tab:c3hp}
  \end{table*}}

\newcommand{\TabHydrocarbonFitResults}{%
  \begin{table*}
    \centering{} %
    \caption{Gaussian fit results for the lines of small hydrocarbons
      detected toward the Horsehead PDR at RA=5h40m53.936s,
      Dec=$-2^\circ28'00''$ 
      (J2000). Only lines in the 3\mm{} band with
      an energy of the upper level lower than 100\K{} and 
      a higher signal-to-noise ratio than 8 are shown.}
    \begin{tabular}{lccccc}
      \hline
      \hline          
      Species & Transition & Frequency & Area & Velocity & Width \\
              &            & MHz & \mKkms{} & \kms{} & \kms{} \\ 
      \hline
      \multirow{6}{*}{\CCH{}} 
      & $N=1-0, J=3/2-1/2~F=1-1$ &  87284.105 &  238$\pm$3   & 10.675$\pm$0.005 & 0.735$\pm$0.013\\
      & $N=1-0, J=3/2-1/2~F=2-1$ &  87316.898 & 1943$\pm$6   & 10.682$\pm$0.001 & 0.815$\pm$0.003\\
      & $N=1-0, J=3/2-1/2~F=1-0$ &  87328.585 & 1035$\pm$4   & 10.675$\pm$0.002 & 0.769$\pm$0.004\\
      & $N=1-0, J=1/2-1/2~F=1-1$ &  87401.989 & 1103$\pm$5   & 10.694$\pm$0.002 & 0.785$\pm$0.004\\
      & $N=1-0, J=1/2-1/2~F=0-1$ &  87407.165 &  489$\pm$4   & 10.697$\pm$0.003 & 0.749$\pm$0.008\\
      & $N=1-0, J=1/2-1/2~F=1-0$ &  87446.470 &  262$\pm$3   & 10.700$\pm$0.004 & 0.745$\pm$0.011\\
      \hline                                                              
      \multirow{4}{*}{\cCCCH{}}  
      & $N=2_{12}-1_{11}, J=5/2-3/2~F=3-2$ &  91494.231 &  106$\pm$3  & 10.395$\pm$0.010 & 0.691$\pm$0.024\\
      & $N=2_{12}-1_{11}, J=5/2-3/2~F=2-1$ &  91497.525 &   59$\pm$3  & 10.561$\pm$0.012 & 0.594$\pm$0.034\\
      & $N=2_{12}-1_{11}, J=3/2-1/2~F=1-0$ &  91692.752 &   32$\pm$3  & 10.712$\pm$0.036 & 0.887$\pm$0.084\\
      & $N=2_{12}-1_{11}, J=3/2-1/2~F=2-1$ &  91699.471 &   74$\pm$4  & 10.760$\pm$0.020 & 0.858$\pm$0.050\\
      \hline                                                              
      \multirow{4}{*}{\lCCCH{}}
      & $J=9/2-7/2~F=5-4(f)$ &  97995.166 &   44$\pm$7  & 10.605$\pm$0.051 & 0.647$\pm$0.108\\
      & $J=9/2-7/2~F=4-3(f)$ &  97995.913 &   45$\pm$10 & 10.583$\pm$0.092 & 0.848$\pm$0.235\\
      & $J=9/2-7/2~F=5-4(e)$ &  98011.611 &   41$\pm$7  & 10.629$\pm$0.046 & 0.599$\pm$0.116\\
      & $J=9/2-7/2~F=4-3(e)$ &  98012.524 &   35$\pm$8  & 10.550$\pm$0.075 & 0.651$\pm$0.183\\
      \hline                                                                 
      \multirow{5}{*}{\cCCCHH{}} 
      & $2_{02}-1_{11}$ &  82093.542 &  242$\pm$6  & 10.704$\pm$0.009 & 0.741$\pm$0.023\\
      & $3_{12}-3_{03}$ &  82966.200 &   68$\pm$3  & 10.660$\pm$0.012 & 0.618$\pm$0.029\\
      & $3_{22}-3_{13}$ &  84727.696 &   14$\pm$2  & 10.633$\pm$0.028 & 0.344$\pm$0.051\\
      & $2_{12}-1_{01}$ &  85338.893 &  748$\pm$4  & 10.687$\pm$0.002 & 0.767$\pm$0.005\\
      & $3_{03}-2_{12}$ & 117151.191 &  507$\pm$72 & 10.680$\pm$0.051 & 0.715$\pm$0.117\\
      \hline                                                                
      \multirow{7}{*}{\lCCCHH{}}
      & $4_{14}-3_{13}$ &  82395.090 & 107$\pm$7 & 10.681$\pm$0.020 & 0.720$\pm$0.057\\ 
      & $4_{04}-3_{03}$ &  83165.256 &  43$\pm$4 & 10.362$\pm$0.029 & 0.647$\pm$0.068\\ 
      & $4_{13}-3_{12}$ &  83933.700 &  93$\pm$3 & 10.690$\pm$0.010 & 0.689$\pm$0.025\\ 
      & $5_{15}-4_{14}$ & 102992.379 &  81$\pm$3 & 10.658$\pm$0.012 & 0.657$\pm$0.030\\
      & $5_{33}-4_{32}$ & 103914.354 &  27$\pm$2 & 10.592$\pm$0.023 & 0.538$\pm$0.052\\
      & $5_{05}-4_{04}$ & 103952.926 &  34$\pm$3 & 10.803$\pm$0.030 & 0.652$\pm$0.068\\
      & $5_{14}-4_{13}$ & 104915.583 &  58$\pm$4 & 10.696$\pm$0.025 & 0.703$\pm$0.058\\
      \hline
      \multirow{8}{*}{\CCCCH{}}
      & $N=9-8, J=19/2-17/2$   &  85634.004 & 198$\pm$4  & 10.737$\pm$0.007 & 0.689$\pm$0.018\\ 
      & $N=9-8, J=17/2-15/2$   &  85672.579 & 168$\pm$4  & 10.725$\pm$0.007 & 0.663$\pm$0.018\\ 
      & $N=10-9, J=21/2-19/2$  &  95150.388 & 204$\pm$3  & 10.734$\pm$0.004 & 0.682$\pm$0.011\\ 
      & $N=10-9, J=19/2-17/2$  &  95188.946 & 177$\pm$3  & 10.700$\pm$0.005 & 0.665$\pm$0.013\\ 
      & $N=11-10, J=23/2-21/2$ & 104666.565 & 202$\pm$4  & 10.732$\pm$0.007 & 0.714$\pm$0.019\\ 
      & $N=11-10, J=21/2-19/2$ & 104705.106 & 177$\pm$5  & 10.705$\pm$0.011 & 0.728$\pm$0.026\\ 
      & $N=12-11, J=25/2-23/2$ & 114182.512 & 162$\pm$12 & 10.665$\pm$0.021 & 0.582$\pm$0.051\\ 
      & $N=12-11, J=23/2-21/2$ & 114221.040 & 186$\pm$12 & 10.663$\pm$0.022 & 0.719$\pm$0.056\\ 
      \hline
    \end{tabular}
    \label{tab:hydro:gaussfit}
  \end{table*}}

\newcommand{\TabChemSD}{%
  \begin{table}
    \centering{} %
    \caption{Column densities and abundances with respect to the
      number of protons, \ie{}, [X] = 0.5 $N$(X)/$N$(\HH{}), toward the PDR
      position, from single-dish observations at resolutions between 25 and 
      $28''$.}
    \begin{tabular}{ccc}
      \hline
      \hline
      Molecule & Column density & Abundance\\
               &    \pscm{}     &          \\
      \hline
      \HH{}      &  $7.2\pm2.4 \times10^{21}$ & 0.5\\
      \CCH{}     &  $1.6\pm0.2 \times10^{14}$ & $1.1\pm0.4 \times 10^{ -8}$\\
      \cCCCH{}   &  $3.9\pm0.5 \times10^{12}$ & $2.7\pm1.0 \times 10^{-10}$\\
      \lCCCH{}   &  $2.1\pm0.7 \times10^{12}$ & $1.4\pm0.7 \times 10^{-10}$\\
      \cCCCHH{}  &  $9.3\pm0.2 \times10^{12}$ & $6.4\pm2.1 \times 10^{-10}$\\
      \lCCCHH{}  &  $2.7\pm0.5 \times10^{12}$ & $1.9\pm1.1 \times 10^{-10}$\\
      \hline
    \end{tabular}
    \label{tab:chem:30m}
  \end{table}}

\newcommand{\TabChemPdBI}{%
  \begin{table*}
    \centering{} %
    \caption{Comparison of the measured and modeled abundances with respect
      to the number of protons at three different positions across the PDR 
      front.}
    {\tiny %
      \begin{tabular}{cccccccc}
        \hline
        \hline
                 & $\delta x$ & \multicolumn{2}{c}{\CCH{}} & \multicolumn{2}{c}{\cCCCHH{}} & \multicolumn{2}{c}{\CCCHp{}} \\
                 & $['']$     & Measured$^{a}$ & Modeled   & Measured$^{a}$ & Modeled      & Measured$^{b}$ & Modeled \\
        \hline
        Cloud    & 24.9 & $(0.6-1.9)\times10^{-9}$ & $1.2\times10^{-10}$ & $(2.2-8.6)\times10^{-11}$ & $1.8\times10^{-11}$ & ---                       & $5.0\times10^{-14}$ \\
        IR peak  & 13.2 & $(0.9-2.7)\times10^{-8}$ & $2.7\times10^{ -9}$ & $(0.5-2.6)\times10^{ -9}$ & $5.7\times10^{-12}$ & $(1.9-4.2)\times10^{-11}$ & $1.0\times10^{-11}$ \\
        IR edge  &  7.4 & $(0.8-3.7)\times10^{-8}$ & $9.7\times10^{-10}$ & $(0.4-3.8)\times10^{ -9}$ & $1.3\times10^{-12}$ & ---                       & $3.7\times10^{-12}$ \\
        \hline
      \end{tabular}}\\
    $^{a}$~\citet{pety05}, deduced from observations at $\sim6''$-resolution.\\
    $^{b}$ This work, using a simple source model to compute the beam dilution.
    \label{tab:chem:pdbi}
  \end{table*}}

\begin{document}

\title{The IRAM-30m line survey of the Horsehead PDR:\\
  II. First detection of the \lCCCHp{} hydrocarbon cation\thanks{Based on
    observations obtained with the IRAM-30m telescope.  IRAM is supported
    by INSU/CNRS (France), MPG (Germany), and IGN (Spain).}}

\titlerunning{First detection of a hydrocarbon cation, \lCCCHp{}, in the
  Horsehead PDR}

\author{J.~Pety\inst{\ref{IRAM},\ref{LERMA}} \and
  P.~Gratier\inst{\ref{IRAM}} \and V.~Guzm\'an\inst{\ref{IRAM}} \and
  E.~Roueff\inst{\ref{LUTH}} \and M.~Gerin\inst{\ref{LERMA}} \and
  J.R.~Goicoechea\inst{\ref{CAB}} \and \\
  S.~Bardeau~\inst{\ref{IRAM}} \and A.~Sievers\inst{\ref{IRAM-SPAIN}} \and
  F.~Le~Petit\inst{\ref{LUTH}} \and J.~Le~Bourlot\inst{\ref{LUTH}} \and
  A.~Belloche\inst{\ref{Bonn}}, D.~Talbi\inst{\ref{LUPM}}}

\authorrunning{Pety, Gratier, Guzm\'an \etal{}}

\institute{%
  IRAM, 300 rue de la Piscine, 38406 Saint Martin d'H\`eres, France\\
  \email{[pety;gratier;guzman;bardeau]@iram.fr}\label{IRAM} %
  \and LERMA, UMR 8112, CNRS and Observatoire de Paris, 61 avenue de
  l'Observatoire, 75014 Paris, France \label{LERMA} \\
  \email{maryvonne.gerin@lra.ens.fr} 
  \and LUTH, UMR 8102, CNRS and Observatoire de Paris, Place J. Janssen,
  92195 Meudon Cedex, France. \label{LUTH} \\
  \email{evelyne.roueff@obspm.fr} %
  \and Centro de Astrobiolog\'{i}a.  CSIC-INTA. Carretera de Ajalvir, Km 4.
  Torrej\'{o}n de Ardoz, 28850 Madrid, Spain \label{CAB} \\
  \email{jr.goicoechea@cab.inta-csic.es} %
  \and{} IRAM, 7 Avenida Pastora, Granada, Spain \label{IRAM-SPAIN} \\
  \email{sievers@iram.es} %
  \and Max-Planck Institut f\"ur Radioastronomie, Auf dem H\"ugel 69, 53121
  Bonn, Germany\label{Bonn}\\
  \email{belloche@mpifr-bonn.mpg.de} %
  \and LUPM, UMR 5299, Université Montpellier 2, Place Eugène Bataillon, 34095 Montpellier cedex 05, France \label{LUPM}\\
  \email{dahbia.talbi@univ-montp2.fr}}

\date{}
\offprints{J.~Pety}%

\abstract %
{Pure gas-phase chemistry models do not succeed in reproducing the measured
  abundances of small hydrocarbons in the interstellar medium.  Information
  on key gas-phase progenitors of these molecules sheds light
  on this problem.} %
{We aim to constrain the chemical content of the Horsehead mane with a
  millimeter unbiased line survey at two positions, namely the
  photo-dissociation region (PDR) and the nearby shielded core.  This
  project revealed a consistent set of eight unidentified lines toward the
  PDR position. We associate them to the \lCCCHp{} hydrocarbon cation,
  which enables us to constrain the chemistry of small hydrocarbons.
  We observed the lowest detectable \J{} line in the millimeter domain
  along a cut toward the illuminating direction to constrain the spatial
  distribution of the \lCCCHp{}
  emission perpendicular to the photo-dissociation front.} %
{We simultaneously fit 1) the rotational and centrifugal distortion
  constants of a linear rotor, and 2) the Gaussian line shapes located at
  the eight predicted frequencies. A rotational diagram is then used to
  infer the excitation temperature and the column density. We finally
  compare the abundance to the results of the Meudon PDR photochemical model.} %
{Six out of the eight unidentified lines observable in the millimeter bands
  are detected with a signal-to-noise ratio from 6 to 19 toward the
  Horsehead PDR, while the two last ones are tentatively detected. Mostly
  noise appears at the same frequency toward the dense core, located less
  than $40''$ away.  Moreover, the spatial distribution of the species
  integrated emission has a shape similar to radical species such as HCO,
  and small hydrocarbons such as \CCH{}, which show enhanced abundances
  toward the PDR.  The observed lines can be accurately fitted with a
  linear rotor model, implying a $^1\Sigma$ ground electronic state.  The
  deduced rotational constant value is $\Be= 11244.9512\pm0.0015\MHz$,
  close to that of \lCCCH{}.} %
{This is the first detection of the \lCCCHp{} hydrocarbon in the
  interstellar medium. Laboratory spectroscopy is underway to confirm these
  results. Interferometric imaging is needed to firmly constrain the small
  hydrocarbon chemistry in the Horsehead.}

   
\maketitle %

\section{Introduction}

Molecular ions play an important role in the physics and chemistry of the
insterstellar medium. They trace the gas physical conditions, \eg{}, its
ionization rate and its ionization fraction \citep[see for
example][]{goicoechea09}, and they participate in the coupling of the gas
with the magnetic field.  Moreover, molecular ions are key species in the
gas phase synthesis of molecules because ion-molecule reactions most often
have no activation barrier.

\FigSpec{} %

Simple hydrocarbon molecules, such as \CCH{}, \CCCH{}, and \CCCHH{} have
been detected in a wide variety of sources from
diffuse~\citep[\eg][]{lucas00} to dark
clouds~\citep[\eg{}][]{wootten80,mangum90}. The high abundances found at
the UV-illuminated edges of molecular clouds or photo-dissociation regions
(PDRs) cannot be reproduced by current pure gas-phase
models~\citep{fuente03,teyssier04}. \citet{pety05} proposed another
chemical route, namely the photo-erosion of Polycyclic Aromatic
Hydrocarbons (PAHs) and small carbon grains to produce these hydrocarbons.
\citet{rimmer12} showed that using a column-dependent cosmic ray ionization
rate slightly improves the agreement between models and observations in the
Horsehead PDR. Ion-molecule reactions with the \CCHp{} and \CCCHp{} cations
are thought to be the most important gas-phase channels to form small
hydrocarbons~\citep{turner00,wakelam10}, but they have not yet been
observed in the interstellar medium. Constraining the abundances of these
intermediate hydrocarbon cations will shed light on the formation routes of
hydrocarbons.

In this paper, we report the first detection in the interstellar medium of
the \lCCCHp{} hydrocarbon cation toward the Horsehead PDR~\citep[more
precisely, at the peak of the HCO emission,][]{gerin09}.
Sect.~\ref{sec:obs} describes the observations.  Sect.~\ref{sec:spectro}
and~\ref{sec:id} explain how we inferred the spectroscopic parameters
associated with the set of unidentified lines, and why we attribute these
to \lCCCHp{}. Sect.~\ref{sec:chem} discusses the determination of the
\lCCCHp{} abundance and its chemistry.

\section{Observations}
\label{sec:obs}

The Horsehead WHISPER project (Wideband High-resolution Iram-30m Surveys at
two Positions with Emir Receivers, PI: J.~Pety) is a complete unbiased line
survey of the 3, 2, and 1\mm{} bands that is being completed at the
IRAM-30m telescope. Two positions are observed: 1) the ``HCO peak''
(RA=5h40m53.936s, Dec=$-2^\circ28'00''$, J2000), which is characteristic of
the photo-dissociation region at the surface of the Horsehead
nebula~\citep{gerin09}, and 2) the ``\DCOp{} peak'' (RA=5h40m55.61s,
Dec=$-2^\circ27'38''$, J2000), which belongs to a cold and shielded
condensation located less than $40''$ away from the PDR edge, where \HCOp{}
is highly deuterated~\citep{pety07}. The combination of the new EMIR
receivers and Fourier transform spectrometers at the IRAM-30m telescope
yields a spectral survey with unprecedented combination of bandwidth
(36\GHz{} at 3\mm{}, 19\GHz{} at 2\mm{}, and 76\GHz{} at 1\mm{}), spectral
resolution (49\kHz{} at 3 and 2\mm; and 195\kHz{} at 1\mm), and sensitivity
(median noise 8.1\mK{}, 18.5\mK{}, and 8.3\mK, respectively). A detailed
presentation of the observing strategy and data reduction process will be
given in another paper. In short, any frequency was observed with two
different frequency tunings and the Horsehead PDR and dense core positions
were alternatively observed every 15 minutes.  The total observing time
amounted to one hour per frequency setup and position.

\TablCCCHp{} %

In the analysis of the survey, we found an 89\mK{} line peak around
89.957\GHz{}, that could not be associated to any transition listed in the
common public catalogs:
CDMS\footnote{\texttt{http://www.astro.uni-koeln.de/cdms/}}~\citep{muller01},
JPL\footnote{\texttt{http://spec.jpl.nasa.gov/}}~\citep{pickett98}, and
splatalogue\footnote{\texttt{http://www.splatalogue.net}}.  The observing
strategy allowed us to rule out that the detected line is a ghost line
incompletely rejected from a strong line in the image side band (the
typical rejection of the EMIR sideband separating mixers is 13\,dB). Our
search for an identification started with the simplest assumption, \ie{},
the associated species is a linear rigid rotor. The frequency of the
transition ($\J+1 \rightarrow \J$) is then given by
\begin{equation}
  \nu = 2 \Be \, (\J+1),
\end{equation}
where \Be{} is the rotational constant. In this simple model, the ratio of
line frequencies depends only on the ratio of $\J+1$ values. We thus could
predict different sets of frequencies to search for companion lines
associated to the same species, each set associating a given $\J+1
\rightarrow \J$ transition to the frequency of the detected unidentified
line, \ie{}, $\sim89.957\GHz{}$. The only set of frequencies that
consistently brings five other detected lines and two more tentative
detections less than 1\MHz{} from the frequency predictions was the one
which associates the 89.957\GHz{} unidentified line to the
$\J=4\rightarrow3$ transition of a linear rigid rotor. Using the \WEEDS{}
extension~\citep{maret11} of the \GILDAS{}/\CLASS{} software\footnote{See
  \texttt{http://www.iram.fr/IRAMFR/GILDAS} for more information about the
  \GILDAS{} softwares.}~\citep{pety05b}, we quickly ruled out the
assignment of any of the detected unidentified lines to other possible
species because the potential candidate species were complex molecules for
which many other expected transitions were not detected in the survey.

Fig.~\ref{fig:c3hp:spec} displays the spectra at the Horsehead PDR and
dense core positions of the consistent set of eight unidentified lines,
which lie in the millimeter frequency bands. Even though the weather
conditions, the pointing and focus corrections, and the tuning setups were
shared for the two observed positions, the unidentified lines were detected
with a profile, integrated signal-to-noise ratio between 3 and 19 at the
PDR position, while only two-sigma upper limits of typically 25\mKkms{} for
a 0.8\kms{} linewidth could be derived at the dense core position. The only
exception is the $\J=4-3$ line, which is tentatively detected toward the
dense core.  Assuming that the emission arises from a Gaussian filament of
$12''$ full width at half maximum centered on the
PDR~\citep[see][]{gerin09}, 27\% of the emission detected at the core
position is explained by beam pickup from the PDR. The remaining emission
could arise in the lower density skin of the dense core, already detected
in HCO~\citep{gerin09} and \CFp{}~\citep{guzman12a,guzman12b}. In summary,
the eight (tentatively) detected lines in the PDR are \emph{unlikely} to be
observing artifacts.

\section{Associated spectroscopic constants}
\label{sec:spectro}

To compute the spectroscopic parameters associated with the set of
unidentified lines, we used higher order improvements to the simple model
of a linear rotor. The second and third order corrections, which include
the effect of the centrifugal distortion, predict that the frequency of the
$\J+1 \rightarrow \J$ transition is
\begin{equation}
  \nu = 2 \Be (\J+1) - 4 \De (\J+1)^3, \quad \mbox{and}
\end{equation}
\begin{equation}
  \nu = 2 \Be (\J+1) - 4 \De (\J+1)^3 + \He (\J+1)^3 [(\J+2)^3 - \J^3],
\end{equation}
where \De{} and \He{} are the centrifugal distortion constants to the
second and third order, respectively. 

Using these expressions, it is possible to generate the sum of 1) Gaussian
white noise, 2) a baseline offset, and 3) Gaussian line profiles at the
eight predicted frequencies of arbitrary areas and line widths. This allows
us to \emph{simultaneously} fit the noise level, a residual baseline
offset, and the Gaussian parameters (areas and line widths) as well as the
rotational and centrifugal distortion constants associated with these
models. In practice, it is very difficult to make the fit converge with the
eight lines because of the limited signal-to-noise ratio of some of them.
We therefore assume that the eight lines have the same line width. This is
likely for two reasons. First, the line width is dominated by the turbulent
velocity field and the lines are probably emitted from the same gas cells
because the range of energy probed by the eight transitions is relatively
narrow (from 10 to 80\K). Second, the velocity gradient in the plane of the
sky is shallow, ensuring that the line width does not evolve significantly
when the beam size changes with frequency. This assumption enables a quick
fit convergence when we start the fit with the following initial values:
$\Be = 11.2\GHz$, $\De=5\kHz$, $\He=1\Hz$, zero offsets, 10\mK{} noise
levels, and $1\kms$ common linewidth.  Appendix~\ref{sec:likelihood}
discusses why the third order model gives the best fit to the data and
Table~\ref{tab:c3hp} gives the results of the fits (both the Gaussian line
parameters and the spectroscopic constants).

Because laboratory measurements are not available, the only unambiguous
frequency available is the one measured in the local standard of rest (LSR)
frame, \Flsr{}. The frequency given in the source frame, \Fsou{}, assumes
an LSR systemic velocity of the source, \vlsr{}. These frequencies are
linked through $\Fsou = \Flsr \,\bracket{1-(\vlsr/\clight)},$ where
\clight{} is the speed of light. The systemic velocity may vary from
species to species because of different coupling between the gas kinematics
and its chemistry and/or line-shape variations due to optical-depth
effects. The analysis of the unambiguously attributed detected lines in the
survey shows that the possible range of LSR systemic velocities in the
Horsehead PDR is quite narrow, $\sim 0.2\kms$ around 10.7\kms{} (see
Fig.~\ref{fig:hydrocarbons:30m} which displays different lines of several
hydrocarbons). All frequencies quoted here accordingly assume an LSR
velocity of 10.7\kms{}. The derived spectroscopic parameters may therefore
need to be slightly linearly scaled.

\FigHydrocarbonsPdBI{} %

\section{Attribution of the lines to the \lCCCHp{} cation}
\label{sec:id}

\FigProfile{} %

The unidentified lines were all detected at the PDR position, but not at
the dense core position. We thus complemented these data with a cut from
the \HII{} region into the molecular cloud along the direction of the
exciting star, $\sigma$~Ori, \ie{}, perpendicular to the photo-dissociation
front.  Figure~\ref{fig:profile} displays the integrated intensity emission
of the $\J=4-3${} unidentified line as a function of the angular distance
from the photo-dissociation front, along with the emission of HCO,
\HthCOp{}, and \CCH{} species (Table~\ref{tab:profile} defines the observed
lines). All these lines were observed simultaneously with the IRAM-30m
telescope during 4 hours of mild summer weather (typically 11\mm{} of
precipitable water vapor), but they all were well-detected nevertheless.
At $29''$ resolution, it is clear that the unidentified species cut peaks
in the UV-illuminated part of the Horsehead mane in the same way as HCO and
\CCH{}, while \HthCOp{} peaks in the dense core, shielded from the UV
field.  This led us to conclude that the species is a reactive molecule
with a spatial distribution similar to small hydrocarbon chains. For
reference, Fig.~\ref{fig:hydrocarbons:pdbi} displays the integrated
intensity emission of the same lines, imaged at $\sim6''$ with the Plateau
de Bure Interferometer~\citep{pety07,gerin09,pety05}, plus the 7.7\mim{}
PAH emission imaged at $\sim6''$ with ISO~\citep{abergel03}.

The quality of the spectroscopic fit suggests that the molecule is a linear
rotor with a $^1\Sigma^+$ electronic ground state, \ie{}, with a closed
electronic shell. The rotational constant is $\Be \sim 11.24\GHz$, which
implies the presence of several heavy atoms in the species.  According to
the literature, the most probable canditate is the \lCCCHp{} cation.
Indeed, ab initio calculation implies that 1) the linear structure is the
most stable, 2) it has the right electronic state, and 3) the computed
rotational constant value is about
11.1\GHz{}~\citep{radom76,wilson80,wilson82,cooper88,ikuta97}.
Experimental spectroscopic confirmation for this cation is being performed
at the PhLAM laboratory in Lille (Bailleux \& Margules, priv.  comm.).

\TabProfile{} %

\citet[]{wilson80} estimated the dipole moment of this cation to be
2.6\Debye{}, but this is a rather old value. We therefore computed it again
with more sophisticated ab initio techniques of quantum chemistry, as
implemented in the MOLPRO suite of programs\footnote{H.-J.~Werner \& P.  J.
  Knowles, MOLPRO (version 2002.6) package of ab initio 254 programs,
  2002.}.  We used the CASSCF-MRCI level of theory with the
correlation-consistent aug-cc-pVQZ basis sets of~\citet{woon93} for all
atoms. The active space of the CASSCF included the $n = 2$ orbitals of
carbon and the 1S orbital of H.  The geometry was first optimized at the
CASSCF level, leading to bond distances for \lCCCHp{} (1.090\AA, 1.246\AA,
1.355\AA) that perfectly match the results obtained in~\citet{ikuta97}.
The dipole moment was then computed at the center of mass of the molecule
for the CASSCF optimized geometry. The resulting dipole moment of \lCCCHp{}
is 3\Debye{} at the CASSCF-MRCI level of theory, \ie{}, close to the
initial value of \citet[]{wilson80}.  We used the new value to compute the
column density of \lCCCHp{}.

\TabChemPdBI{} %

\section{\lCCCHp{} abundance and chemistry}
\label{sec:chem}

\FigRotDiag{} %

We used the formalism of the rotational diagram to estimate the column
density of \lCCCHp{}. \citet{goldsmith99} give the expressions for the
level energies, the quantum level degeneracies, the Einstein coefficients,
and the partition function $Z(T)$ for a rigid rotor of dipole moment,
$\mu$. For instance, the Einstein coefficients are given by
\begin{equation}
  \A{\J+1}{\J} = 
  \frac{64 \pi^{4} \nu^{3} \mu^{2}}{3\,h\,c^3} \frac{\J+1}{2\J+3}, 
\end{equation}
where $h$ and $c$ are the Planck constant and the light speed,
respectively. Using these formula, we are able to compute the rotational
diagram, assuming 1) that the lines are optically thin $(\tau< 1)$, and 2)
that the \lCCCHp{} emission is co-spatial with the illuminated filament of
the \CCH{} emission, \ie{}, it approximately fills as a Gaussian filament
of $\sim 12''$ width in the $\delta x$ direction, infinite size in the
$\delta y$ direction, and centered at the HCO peak (see
Fig.~\ref{fig:hydrocarbons:pdbi}). The filling factors are 0.4, 0.6, and
0.8 at 90, 157, and 270\GHz{}, respectively.  Figure~\ref{fig:rot:diag}
displays the diagram for $\mu_0=3\Debye$ (this work). The rotational
temperature inferred from the observations is independent of the dipole
moment because it scales with the slope of the rotational diagram, while
the column density scales as $\mu^{-2}$.  The typical rotational
temperature and column density thus are $14\pm{}2\K$ and
$(\mu_0/\mu)^2\,(4.8\pm0.9)\times10^{11}\pscm,$ respectively.

We took advantage of the Horsehead WHISPER survey to make a consistent
summary of the small hydrocarbon detections in the Horsehead PDR. The
spectra of the detected lines in the 3\mm{} band of \CCH{}, \cCCCH{},
\lCCCH{}, \cCCCHH{}, \lCCCHH{}, and \CCCCH{} are displayed in
Fig.~\ref{fig:hydrocarbons:30m} and their Gaussian fit results are
summarized in Table.~\ref{tab:hydro:gaussfit} (both available on-line
only). The integrated intensities of the lines are consistent with the
results published by \citet{teyssier04,teyssier05}. We accordingly just
summarized their column densities and abundances in
Table~\ref{tab:chem:30m}.  \citet{pety05} presented higher angular
(typically $6''$) observations of \CCH{} and \cCCCHH{} obtained with the
Plateau de Bure Interferometer. The associated abundances at three
different positions (the ``IR peak'' close to the usual ``HCO peak'', one
position named ``cloud'' representative of the UV-shielded material, and
one position named ``IR edge'' closer to the \HII{} region) are summarized
in Table~\ref{tab:chem:pdbi}.  The comparison of Table~\ref{tab:chem:pdbi}
and~\ref{tab:chem:30m} indicates that the \CCH{} and \cCCCHH{} abundances
measured at a typical resolution of $25-28''$ falls in the uncertainty
range of abundances measured at a four times better resolution with PdBI.
This means that their emission more or less fills the $25-28''$ beam, even
though it is substructured into filaments (see
Fig.~\ref{fig:hydrocarbons:pdbi}). Indeed, if we had assumed that the
\lCCCHp{} emission fills the beam at every measured frequency, the inferred
\lCCCHp{} abundance would have only be reduced by a factor 2.

\TabChemSD{} %
\FigChemistryLog{} %

We finally compared the derived abundances with a one-dimensional,
steady-state photochemical model \citep[Meudon PDR
model,][]{lebourlot12,lepetit06}. The used version of the Meudon PDR code
includes the Langmuir-Hinshelwood and Eley-Rideal mechanisms to describe
the formation of \HH{} on grains~\citep{lebourlot12}, and surface reactions
for other species (Le Bourlot et al., in prep.). The physical conditions in
the Horsehead have already been constrained by our previous observational
studies and we kept the same assumptions for the steep density gradient,
radiation field~\citep[$\chi = 60$ in Draine units, see][]{abergel03},
elemental gas-phase abundances \citep[see Table 6 in][]{goicoechea06} and
cosmic ray primary ionization rate~\citep[$\zeta= 5\times10^{-17}\ps$ per
\HH{} molecule,][]{goicoechea09}.

We used the Ohio State University (osu) pure gas-phase chemical network
upgraded for photochemical studies. \CCCHp{} is produced by reactions
between \CCHH{} and \Cp{}.  Then, \CCCHp{} is thought to produce \CCCHHp{},
and \CCCHHHp{} through reactions with \HH{}, which later recombines with
electrons to form \CCCH{} and \CCCHH{}
\begin{equation*} 
  \xymatrixrowsep{-0.1cm}
  \xymatrixcolsep{0.6cm}
  \xymatrix{
  & &  & \chem{C_3H^+_2} \ar[r]^{\tiny \chem{e^{-}}} &  \CCCH\\
  & \CCHH \ar[r]^{\tiny \Cp}
  & \CCCHp \ar[ru]^{\tiny \HH} \ar[rd]_{\tiny \HH} & & \\
  & &  & \CCCHHHp \ar[r]^{\tiny \chem{e^{-}}}
  & \CCCHH.}
\end{equation*}
Hence these species are usually included in gas-phase chemical networks
with reaction rate accuracies of a factor 2 or better. We can thus use them
to compare models with observations.

Figure~\ref{fig:chemistry:log} presents the results of the photochemical
model for a few small hydrocarbon molecules and cations, namely \CCH{},
\CCCH{}, \CCCHH{}, \CCHp{}, and \CCCHp{}. From top to bottom, the figure
shows the spatial profiles of the density, abundances relative to the
number of hydrogen atoms, and the abundances relative to \CCCHp{}. The left
column presents the profiles computed by the code, which samples the
UV-illuminated gas on a finer spatial grid than the UV-shielded gas to
correctly represent the steep physical and chemical gradients. The right
column presents the profiles convolved with a Gaussian of $6''$ full width
at half maximum to facilitate the comparison with the abundances inferred
from PdBI observations at $6''$ angular resolution.  The measured
abundances are displayed with vertical segments for \CCH{}, and \cCCCHH{},
and with a filled rectangle for \CCCHp{}.  Although the \lCCCHp{} abundance
is only inferred from the IRAM-30m single-dish telescope, we also show it
here because we used a simple model of its emission based on the PdBI
hydrocarbon data to correct for the beam dilution.
Table~\ref{tab:chem:pdbi} quantitatively compares the measured and modeled
abundances at the PDR position (``IR peak'' at $\Av\sim1\magn$), inside the
molecular cloud at a position representative of the UV-shielded material
``cloud'' $(\Av\sim8\magn)$, and closer to the \HII{} region (\ie{}, ``IR
edge'' at $\Av\sim0.01\magn$).

We only consider the most stable isomers to compare the measured and
modeled abundances. Table~\ref{tab:chem:30m} indicates that the amount of
\CCCHH{} locked in the linear species is negligible, and that twice as much
\CCCH{} is locked in the cyclic species compared to the linear species.
Moreover, Table~1 of \citet{mebel07} indicates that the cyclic form of
\CCCHH{} and \CCCH{} are more stable than their linear form by
59.3\,kJ\,mol$^{-1}$ (or 0.61\,eV), and 11\,kJ\,mol$^{-1}$ (or 0.11\,eV),
respectively. Finally, \citet{savic05a} indicated that the cyclic form of
\CCCHp{} is less stable than the linear one by 220\,kJ\,mol$^{-1}$, or
2.28\,eV. Given the large difference in energies, we here assume that the
linear form is the main product in the gas phase.

While \CCH{} has much brighter lines and a column density higher by two
orders of magnitude compared to linear and cyclic \CCCH{}, we have no hint
of a detection of bright lines associated to \CCHp{}. This is due to a
combination of two effects. First, the spectroscopic structure of \CCHp{}
(which has a $^3\Pi$ ground electronic state) is more complex than that of
\CCCHp{}, implying that the emission is spread over more lines.  The
knowledge of the rotational spectrum of \CCHp{} would enable a deep search
of this reactive molecular ion to check whether \CCHp{} is also present in
the horsehead PDR. Second, \CCHp{} is predicted to be less abundant than
\CCCHp{} (except when $\Av\le0.1\magn$), because it is efficiently
destroyed by \HH{}, independently of the temperature. In contrast, the
detection of \CCCHp{} at the warm PDR position corroborates that the
destruction of \CCCHp{} by \HH{} is much less efficient there than in the
cold-core region, \ie{}, the $\CCCHp + \HH$ reaction is strongly dependent
on the gas temperature~\citep{savic05b}.

On one hand, the agreement between the interferometric abundances of \CCH{}
and \CCCHH{} and the modeled curves improved at the ``cloud'' position by
up to four orders of magnitude compared to the models of \citet[][see their
Fig.~10]{pety05}. This is the result of the recent addition of the
chemistry on the grain surface to the Meudon PDR code. Indeed, the modeled
and predicted abundances of \cCCCHH{} are consistent, while the modeled
abundance of \CCH{} is still one order of magnitude lower than the measured
abundance (\cf{} Table~\ref{tab:chem:pdbi}). However, it is worth noting
that the ``cloud'' position ($\delta x = 24.9'',\delta y = -5.3''$) does
not fall, for historical reasons, on the dense core.  Indeed, the \CCH{}
emission presents a dip at the \DCOp{} peak ($\delta x =44.7'', \delta y =
16.5''$, blue cross on Fig.~\ref{fig:hydrocarbons:pdbi}), while the
\cCCCHH{} emission shows a bright filament there. These observational facts
indicate that \CCH{} probably depletes at a higher rate onto grains than
\CCCHH{}.

On the other hand, the model results did not change significantly in the
UV-illuminated part (IR peak and IR edge). The measured abundances are
typically one and two orders of magnitude higher than the modeled
abundances for \CCH{} and \CCCHH{}. In contrast, the measured abundance of
\CCCHp{} is only about a factor 2 higher than the modeled abundance (\cf{}
Table~\ref{tab:chem:pdbi}). If the assumptions used here are correct (the
geometry of the \lCCCHp{} emission, and neglecting the amount of
c-\CCCHp{}), this difference between the small hydrocarbon molecules and
\lCCCHp{} would confirm that there is a non gas-phase chemical way in which
small hydrocarbons are formed in UV-illuminated regions, \eg{}, the
photo-erosion of PAHs~\citep{pety05}.  Interferometric observations of
\CCCH{} and \CCCHp{} are needed to confirm this result. Indeed,
Fig.~\ref{fig:chemistry:lin} shows the spatial profile of the modeled
\CCCHp{} abundance convolved by convolution at $6''$ and $27''$. Clearly,
only interferometric observations can provide the high angular resolution
needed to resolve the actual structure of the \CCCHp{} emission, and
potentially detect a spatial shift with respect to the millimeter wave
emission of other small hydrocarbons and to the infrared imaging of PAHs
and dust grains.

\FigChemistryLin{} %

\section{Summary}

We reported the first detection of \lCCCHp{} in the interstellar medium.
Laboratory measurements of the \lCCCHp{} spectroscopy are needed to improve
the spectroscopic characterization of this molecular ion. Interferometric
imaging with either PdBI or ALMA is required to better constrain the small
hydrocarbon chemistry.

\begin{acknowledgements}
  This work was funded by grant ANR-09-BLAN-0231-01 from the French {\it
    Agence Nationale de la Recherche} as part of the SCHISM project.  VG
  thanks the Chilean Government for support through the Becas Chile
  scholarship program. JRG thanks the Spanish MICINN for his support
  through a Ram\'on y Cajal research contract and additional funding
  through grants AYA2009-07304 and CSD2009-00038. We thank H. S. Liszt for
  finding the neat acronym of this project, ``Horsehead WHISPER''.
\end{acknowledgements}

\bibliographystyle{aa}
\bibliography{aa20062}

%%%%%%%%%%%%%%%%%%%%%%%%%%%%%%%%%%%%%%%%%%%%%%%%%%%%%%%%%%%%%%%%%%%%%%%%%%%

\appendix{}

\section{Fit likelihood}
\label{sec:likelihood}

We wish to know to which development order we can significantly fit the
frequency model to the observed data. Assuming that all intensity channels
are independent measures with Gaussian noise, the likelihood, \likelihood,
of the fit of the model to the data is given by
\begin{equation}
  \likelihood = 
  \prod_{i=1}^{N} \bracket{\frac{1}{\sqrt{2\pi\sigma_i}} \exp\paren{-\frac{[I^\emr{obs}_i-I^\emr{mod}_i(\theta)]^2}{2\sigma_i^2}}},
\end{equation}
where $i$ is the index over the $N$ measured channel intensities,
$\sigma_i$ the noise of channel $i$, $I^\emr{obs}$ and $I^\emr{mod}$ the
observed and modeled intensities, and $\theta$ the set of model parameters.
Taking the logarithm yields $\ln(\likelihood) = K-0.5\chi^{2}$, where
\begin{equation*}
  K = -\frac{1}{2}N\ln(2 \pi\sigma_{i})
  \quad \mbox{and} \quad 
  \chi^{2} = \sum_{i=1}^{N}\frac{[I^\emr{obs}_{i}-I^\emr{mod}_{i}(\theta)]^2}{\sigma_{i}^2}.
\end{equation*}
$K$ is a constant and $\chi^{2}$ is the usual chi square measure of the fit
quality. If we try to fit two models $M_{1}$ and $M_{2}$, which differs
only by one parameter, to the same data, the comparison of the fits is said
to have only one degree of freedom. The difference
$\chi^2_{M_{1}}-\chi^2_{M_{2}} =
2\ln\paren{\likelihood_{M_{2}}/\likelihood_{M_{1}}}$ then follows the well
known $\chi^{2}$ distribution with one degree of freedom. In our case, the
difference of $\chi^2$ between the second and third order models is 9.2,
which implies that the third-order model better reproduces the data than
the second-order model with a probability $\ge 99.75\%$. In contrast,
developing the centrifugal effect to the fourth order does not bring a
significant improvement over the third order development. We therefore
stopped the determination of the spectroscopic constants to the third order
model.

\Online{}

\FigHydrocarbonsSD{} %
\TabHydrocarbonFitResults{} %

\end{document}